\documentclass[
  twocolumn,
  prb,
  showpacs,
  amsmath,
  amssymb,
  superscriptaddress,
  floatfix
]{revtex4}

\usepackage{bm}
\usepackage{graphicx}

\def\be{\begin{equation}}
\def\ee{\end{equation}}
\def\bea{\begin{eqnarray}}
\def\eea{\end{eqnarray}}
\def\ve{\varepsilon}

\def\tq{{\tau_{\rm q}}}
\def\ttr{{\tau_{\rm tr}}}

\def\wc{{\omega_c}}
\def\ve{\varepsilon}

\def\w{\omega}

\def\be{\begin{equation}}
\def\ee{\end{equation}}
\def\bea{\begin{eqnarray}}
\def\eea{\end{eqnarray}}
\def\ve{\varepsilon}
\def\wc{\omega_c}
\def\w{\omega}
\def\ve{\varepsilon}
\def\w{\omega}

\def\tq{\tau_{\rm q}}

\def\ttr{\tau_{\rm tr}}

\begin{document}

\title{Quantum oscillations in the microwave magnetoabsorption of a 2D electron gas.}
\author{O. M.~Fedorych}
  \affiliation{Grenoble High Magnetic Field Laboratory, CNRS,
  Grenoble, France}
\author{M.~Potemski}
  \affiliation{Grenoble High Magnetic Field Laboratory, CNRS,
  Grenoble, France}
\date{\today}
\author{S. A.~Studenikin}
  \affiliation{Institute for Microstructural Sciences, NRC, Ottawa,
  Ontario K1A-0R6, Canada}
\author{J. A.~Gupta}
  \affiliation{Institute for Microstructural Sciences, NRC, Ottawa,
  Ontario K1A-0R6, Canada}
\author{Z. R.~Wasilewski}
  \affiliation{Institute for Microstructural Sciences, NRC, Ottawa,
  Ontario K1A-0R6, Canada}
\author{I. A. Dmitriev}\thanks{Also at Ioffe Physical Technical Institute, 194021 St.~Petersburg, Russia}
  \affiliation{ Institute of Nanotechnology, Karlsruhe Institute of Technology, 76021 Karlsruhe, Germany}

%
\begin{abstract}
We report on the experimental observation of the quantum oscillations in microwave magnetoabsorption of a high-mobility
two-dimensional electron gas induced by Landau quantization. Using original resonance-cavity technique, we observe two kinds of
oscillations in the magnetoabsorption originating from inter-Landau-level  and intra-Landau-level transitions.  The experimental
observations are in full accordance with theoretical predictions. Presented theory also explains why similar quantum oscillations are
not observed in transmission and reflection experiments on high-mobility structures despite of very strong effect of microwaves on
the dc resistance in the same samples.
\end{abstract}
 \maketitle

\noindent Quantum oscillations in absorption (QMA) by a two-dimensional electron gas (2DEG) in perpendicular magnetic 
field $B$, governed by the ratio $\w/\wc$
of the wave frequency $\w=2\pi f_{\rm mw}$ of external electro-magnetic wave and the cyclotron frequency $\wc=e B/m c$, 
were predicted long ago by Ando\cite{Ando} and observed experimentally\cite{Abstreiter} in the IR-absorption on a
low-mobility and high-density 2DEG in a Si-inversion layer. Recently, similar $\w/\wc$--oscillations were discovered in
the dc resistance of a high-mobility 2DEG irradiated by microwaves.\cite{zudov01}  Particularly intriguing are zero-resistance
states\cite{Mani02} which develop in the minima of these microwave-induced resistance oscillations (MIRO).

Theoretically, both MIRO (Refs.~\onlinecite{DMP03,VA,DVAMP05}) and QMA (Refs.~\onlinecite{Ando,DMP03,VA,DVAMP05,Raichev}) stem from
microwave-assisted transitions between disorder-broadened Landau levels (LLs). However, in experiments on
high-mobility samples no QMA were observed so far despite strong MIRO showing up in the same experimental conditions. Several attempts
to measure microwave reflection or transmission simultaneously with MIRO reported either single cyclotron resonance (CR)
peak\cite{Smet,Stud1,Du} or more complex structure dominated by confined magnetoplasmons (CMP).\cite{IEEE,wirthmann07,ourPRB}

Using original resonance-cavity technique, in this work we observe well-pronounced QMA in a high-mobility GaAs/AlGaAs sample which
also reveals strong MIRO in dc transport measurements. For both  the $T$-independent QMA and dynamic Shubnikov-de Haas oscillations (SdHO), 
the experimental results
fully agree with the theoretical predictions of Ref.~\onlinecite{DMP03}, which generalizes theory\cite{Ando} of Ando for the case of
smooth disorder potential appropriate for high-mobility structures. In addition, we  explain the failure to
observe such quantum oscillations in transmission and reflection experiments.

We start with a summary of relevant theoretical results which includes QMA theory\cite{DMP03} for dynamic conductivity at high LLs 
and nonlinear relation between the absorption and dynamic conductivity\cite{quinn76,Stud1,Raichev} specific
for high-mobility 2DEG samples. Consider a plane wave normally incident to the 2DEG at the interface $z=0$ between two dielectrics
with permittivity $\epsilon_1$ ($z<0$) and $\epsilon_2$ ($z>0$). The electric field ${\rm Re}{\bf E}_l$ of external ($l=e$),
reflected ($l=r$), and transmitted ($l=t$) waves is a real part of ${\bf E}_l={\cal E}_l\exp(i k_l z-i \w t)\sum_\pm s^{(l)}_\pm {\bf
e}_\pm $, where the wave numbers $k_e/\sqrt{\epsilon_1}=-k_r/\sqrt{\epsilon_1}=k_t/\sqrt{\epsilon_2}=\w/c$, coefficients
$s^{(l)}_\pm$ describe the polarization,
 $\sum_\pm |s^{(l)}_\pm|^2=1$, and $\sqrt{2}{\bf e}_\pm={\bf e}_x\pm i{\bf e}_y$.
According to the Maxwell equations, boundary conditions at $z=0$ read ${\bf E}_t={\bf E}_r+{\bf E}_e$ and
 $\partial_z({\bf E}_t-{\bf E}_r-{\bf E}_e)=(4\pi/c^2)\,\partial_t\, \hat{\sigma}{\bf E}_t$.
It follows that 
\be\label{t} \sqrt{\epsilon_1}{\cal E}_e s^{(e)}_\pm/{\cal E}_t s^{(t)}_\pm=\sqrt{\epsilon_{\rm eff}}
+2\pi\sigma_\pm/c, \ee
where $2 \sqrt{\epsilon_{\rm eff}}=\sqrt{\epsilon_1}+\sqrt{\epsilon_2}$.
Further, $\sigma_\pm=\sigma_{xx}\pm i\sigma_{yx}$ are the
eigenvalues of the complex conductivity tensor $\hat{\sigma}$ having the symmetries $\sigma_{xx}=\sigma_{yy}$ and
$\sigma_{xy}=-\sigma_{yx}$, namely, $\hat{\sigma}{\bf e}_\pm=\sigma_\pm{\bf e}_\pm$.

Equation (\ref{t}) yields the absorption ${\cal A}$, transmission ${\cal T}$, and reflection ${\cal R}$ coefficients (see also
Refs.~\onlinecite{quinn76,Stud1,Raichev}),
\bea \label{A}
&&{\cal A}=\sum\limits_\pm\frac{\sqrt{\epsilon_1}\,|s^{(e)}_\pm|^2}{|\sqrt{\epsilon_{\rm eff}}+2\pi\sigma_\pm/c|^2}{\rm Re}\frac{4\pi\sigma_\pm}{c}\,,\\
\label{T}
&&{\cal T}=\sum\limits_\pm\frac{\sqrt{\epsilon_1 \epsilon_2}\,|s^{(e)}_\pm|^2}{|\sqrt{\epsilon_{\rm eff}}+2\pi\sigma_\pm/c|^2}\,,\\
\label{R} &&{\cal R}=\sum\limits_\pm|s^{(e)}_\pm|^2\left|\frac{\sqrt{\epsilon_1}-\sqrt{\epsilon_2}-4\pi\sigma_\pm/c}
{\sqrt{\epsilon_1}+\sqrt{\epsilon_2}+4\pi\sigma_\pm/c}\right|^2\,. \eea It is important to mention that the dynamic conductivity
$\sigma_\pm$, which is the focus of present study, is the response to the screened electric field acting on 2D electrons. By
contrast, coefficients (\ref{A})--(\ref{R}) represent a response to the (unscreened) electric component of incoming wave and,
therefore, measure both  single-particle (transport) and collective (screening) properties of 2DEG.

The dynamical screening, represented by the denominators in Eqs.~(\ref{A})--(\ref{R}), becomes particularly strong in high-mobility
structures where the ratio $|2\pi\sigma_\pm/c|$ reaches values much larger than unity. Indeed, in the absence of Landau quantization
the conductivity $\sigma_\pm=\sigma_\pm^D$ is given by the Drude formula,
\be\label{sD} \sigma_\pm^D=\frac{n e^2/m}{\ttr^{-1}-i(\w\pm\wc)}\,, \ee
where $\ttr$ is the momentum relaxation time. The absorption, Eq.~(\ref{A}), takes the form
\be \label{AD}{\cal A}^D=\sqrt{\frac{\epsilon_1}{\epsilon_{\rm eff}}}\sum\limits_\pm|s^{(e)}_\pm|^2\frac{\Omega\tau_{\rm
tr}^{-1}}{(\Omega+\tau_{\rm tr}^{-1})^2+(\w\pm\wc)^2}\,, \ee
where $\hbar\Omega=2\alpha\ve_F/\sqrt{\epsilon_{\rm eff}}$, $\alpha=e^2/\hbar c\simeq1/137$ is the fine structure constant, and
$\ve_F$ is the Fermi energy of 2DEG. In high-mobility 2DEG $\Omega\ttr\gg 1$ and the width of the cyclotron peak in
Eqs.~(\ref{A})--(\ref{R}) and (\ref{AD}) is dominated by strong reflection of microwaves. In the region $|\w-\wc|\lesssim\Omega$,
where $|2\pi\sigma_-|\gg c$, the collective effects are pronounced. In this region, a special care should be taken to avoid the
finite-size magnetoplasmon effects.\cite{IEEE,wirthmann07,ourPRB}

According to Ref.~\onlinecite{DMP03} (which generalizes the results of Ref.~\onlinecite{Ando} for the relevant case of smooth
disorder potential, see also Refs.~\onlinecite{VA} and \onlinecite{Raichev}), Landau quantization at high LLs  modifies Drude formula
to the form which we call the quantum Drude formula (QDF) in what  follows,
\bea \label{QDF} {\rm Re}\sigma_\pm=\frac{n e^2}{\omega m}\!\int\! \frac{d\ve\,(f_\ve-f_{\ve+\omega})\,\tilde{\nu}(\ve)\,\tau_{\rm
tr, B}^{-1}(\ve+\omega)}{[\tau^{-2}_{\rm tr, B}(\ve)+\tau^{-2}_{\rm tr, B}(\ve+\omega)]/2+(\omega\pm\omega_c)^2}~, \eea
 where $f_\ve$ is the Fermi distribution function. The Landau quantization leads to the oscillatory density of states (DOS),
$\nu(\ve)=\nu(\ve+\wc)\equiv\nu_0\tilde{\nu}(\ve)$, and to renormalization of the transport relaxation time, $\tau_{\rm
tr,B}(\ve)\equiv\tau_{\rm tr}/\tilde{\nu}(\ve)$, where $\nu_0=m/2\pi\hbar^2$ is the zero-$B$ DOS per spin orientation. In the limit
of strongly overlapping LLs, $\wc\tq\ll 1$, where $\tau_{\rm q}$ is the quantum relaxation time, the DOS is weakly modulated by magnetic field,
\be\label{OvLLs} \tilde{\nu}(\ve)=1-2\delta\cos\frac{2\pi\ve}{\wc}~,\qquad \delta=e^{-\pi/\wc\tq}\ll 1~.
\ee
In the opposite limit $\wc\tq\gg1$, LLs become separated, $\tilde{\nu}(\ve)=\tq{\rm Re}\sqrt{\Gamma^2-(\ve-\ve_n)^2}$, where
$\Gamma=\sqrt{{2\omega_c}/{\pi\tq}}<\wc/2$~, and $\ve_n$ marks the position of the nearest LL.

At the CR $\w=\wc$, QDF (\ref{QDF}) reads
 \be\label{res} \sigma_-|_{\wc=\w}=\sigma^D_-|_{\wc=\w}\!\int\!d\ve\Theta[\nu(\ve)]\frac{f_\ve-f_{\ve+\w}}{\w}~,
 \ee
 where the integration is over the regions with $\nu(\ve)> 0$. In the case of overlapping LLs, this produces the classical Drude result $\sigma_-=\sigma^D_-$ meaning that quantum effects in the vicinity of the resonance are absent. In the case of separated LLs, $\wc\gg \Gamma$, the conductivity is reduced,
$\sigma_-|_{\wc=\w}=(2\Gamma/\wc)\sigma^D_-$. At the same time, the CR width increases from $\tau_{\rm tr}^{-1}$ to $\tau_{\rm tr}^{-1}\wc/\Gamma$.\cite{DMP03,BIS}

In what follows we consider the region $|\w-\wc|\gg \tau_{\rm tr}^{-1}$ where one can safely neglect $\tau_{\rm tr}^{-1}$ in the
denominator of Eq. (\ref{AD}) and QDF, Eq. (\ref{QDF}), which gives
\be\label{main} {\cal A}/{\cal A}^D=\int\!d\ve\frac{f_\ve-f_{\ve+\w}}{\w}\tilde{\nu}(\ve)\tilde{\nu}(\ve+\w). \ee
Equation (\ref{main}) is the key theoretical result for our study. It expresses the imbalance between the rates of absorption and emission of
the microwave quanta, both proportional to the product of initial and final density of states.
 The screening properties of a 2DEG at $|\w\pm\wc|\gg \tau_{\rm tr}^{-1}$ are solely determined by the
 non-dissipative part of conductivity, ${\rm Im}\sigma_{\pm}\simeq n e^2/m(\w\pm\wc)\gg {\rm Re}\sigma_{\pm}$,
which remains finite at $\tau_{\rm tr}^{-1}\to 0$.  That is why the transmission and reflection coefficients, Eqs. (\ref{T}) and
(\ref{R}), in high-mobility structures are  almost independent of disorder scattering making it very difficult to observe QMA in
direct transmission and reflection experiments.\cite{Smet,Stud1,IEEE,Du,wirthmann07}

\begin{figure}
\includegraphics[width=0.95\columnwidth]{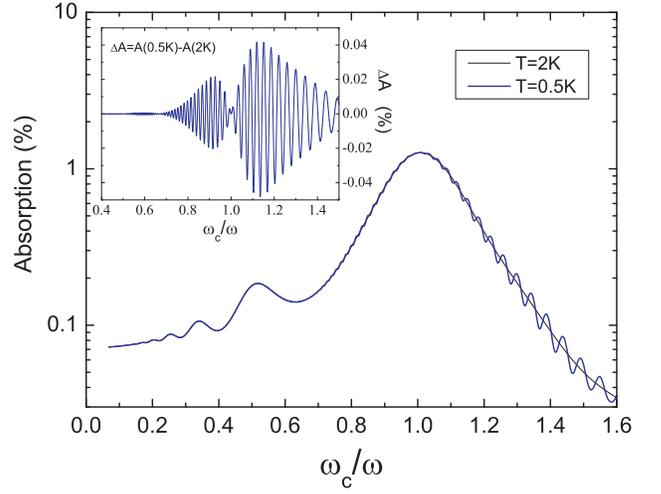}
\caption{Magnetoabsorption calculated using Eqs.~(\ref{main}) and (\ref{AD}) for two temperatures $T=$2~K and 0.5~K. Frequency
$f_{\rm mw}=$60 GHz, density $n=1.8\cdot 10^{11} {\rm cm}^{-2}$, mobility $\mu=10^7 {\rm cm}^2/{\rm V s}$, and $\tau_{\rm q}=$15~ps.
The difference between traces at T=0.5~K and 2~K in the inset shows the $T$-dependent dynamic SdHO.}
 \label{Fig1}
\end{figure}

Magnetoabsorption at two temperatures $T=$2~K and 0.5~K as given by Eqs.(\ref{AD}) and (\ref{main}) is illustrated in
Fig.~\ref{Fig1} for high-mobility 2DEG similar to best structures used for MIRO measurements. The broadening of the envelope Drude
peak (\ref{AD}) is dominated by strong reflection of microwaves near the CR since in our example $\Omega\sim\omega\gg\tau_{\rm
tr}^{-1}$. The Drude peak in Fig.~\ref{Fig1} is modulated by the dynamic SdHO (at $\wc\gtrsim\w$ and $T=$0.5~K) and by the
temperature-independent QMA with maxima at the CR harmonics $\w=N\wc$, which we discuss below.

At low temperature, $X_T=2\pi^2 T/\hbar\wc\lesssim 1$, the absorption (\ref{main}) manifests the dynamic SdHO, which in the case of
overlapping LLs (\ref{OvLLs}) are given by
\be\label{SdH} {\cal A}/{\cal A}^D=1-\frac{4\,X_T \,\delta }{\sinh
X_T}\,\frac{\wc}{2\pi\w}\sin\frac{2\pi\w}{\wc}\cos\frac{2\pi\ve}{\wc}. \ee
The dynamic SdHO are exponentially suppressed at $X_T\gg 1$ similar to SdHO in the dc resistance.  In the  inset it is clearly seen
that the dynamic SdHO are periodically modulated according to $\sin(2\pi\w/\wc)$ with nodes at the CR harmonics.

At $e^{-X_T}\ll\delta$, i.e. $T\gg T_D=\hbar/2\pi\tq$, SdHO (\ref{SdH}) become  exponentially smaller than
$T$-independent $\omega/\omega_c$-oscillations of second order ${\cal O}(\delta^2)$ which represent QMA,
\be\label{mainOvLLs} {\cal A}/{\cal A}^D\simeq\langle\tilde{\nu}(\ve)\tilde{\nu}(\ve+\w)\rangle=2\delta^2\cos\frac{2\pi\w}{\wc}. \ee
Here the angular brackets denote $\ve$--averaging over the period $\omega_c$. Maxima of QMA seen in Fig.~\ref{Fig1} appear at integer
harmonics of the CR $\w/\wc = N$. The amplitude of QMA (\ref{mainOvLLs}) becomes of order unity when the DOS modulation is
pronounced, i.e. $\delta\sim 1$.\cite{memory}

Since QMA lie in the microscopic origin of MIRO \cite{DMP03}, it is instructive to compare the expression (\ref{mainOvLLs}) to
results for MIRO in the same regime. For inelastic mechanism of MIRO \cite{DMP03,DVAMP05}, the photoresistivity $\rho_{\rm ph}$ in
terms of Drude resistivity $\rho^D$ reads
\be\label{ph} \frac{\rho_{\rm ph}}{\rho^D}=1+2\delta^2-\frac{4\tau_{\rm in}{\cal
P}^D}{\omega^2\nu_0}\delta^2\frac{2\pi\omega}{\omega_c}\sin\frac{2\pi\omega}{\omega_c}\,. \ee
where  the microwave power dissipated in the absence of Landau quantization is ${\cal P}^D={\cal A}^D\sqrt{\epsilon_1}c E^2_e/4\pi$,
and $\tau_{\rm in}$ is the inelastic relaxation time. Below we use Eq.~(\ref{ph}) to determine $\tau_{\rm q}$ from dc measurement of
MIRO. The known value of $\tau_{\rm q}$ enables a direct comparison of the measured and calculated QMA
[Eq.~(\ref{mainOvLLs})] without fitting parameters.

{\it Experiment.} Two rectangular samples were cleaved from a single MBE-grown wafer of a high-mobility GaAs/AlGaAs heterostructure
V0050. After illumination with red light the electron concentration was $n=3.6\times 10^{11} {\rm cm}^{-2}$ and mobility $\mu=5\times
10^6 {\rm cm^2/V s}$. Sample~1 was prepared for dc transport measurements. In-Sn Ohmic contacts were made by rapid annealing in
reducing atmosphere of argon bubbled through hydrochloric acid. The sample was placed in a helium cryostat equipped with a
superconducting magnet. Similar to Refs. \onlinecite{Stud1} and \onlinecite{IEEE}, microwaves from an HP source, model E8257D, were
delivered to the sample using Cu-Be coaxial cable terminated with a 3 mm antenna.
 \begin{figure}
\includegraphics[width=0.95\columnwidth]{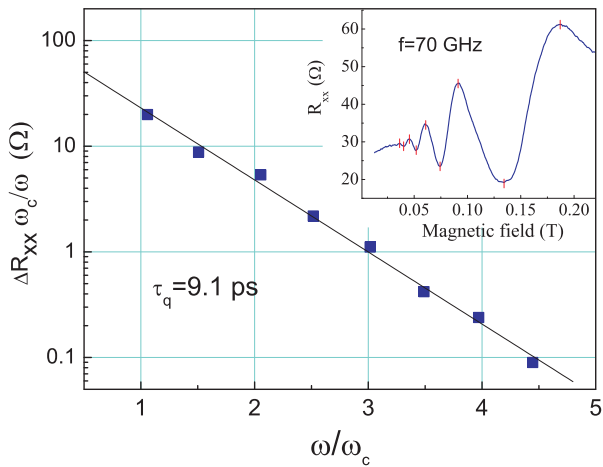}
\caption{MIRO measured on Sample~1 for $f_{\rm mw}=$67 GHz at T=2K (inset), and the amplitude of MIRO vs. $\omega/\omega_c$ in
semi-log scale (main panel).  The linear fit yields the quantum scattering time  $\tau_q =9.1$~ps, see Eqs.~(\ref{ph}) and
(\ref{OvLLs}) .
 }
 \label{Fig2}
\end{figure}

Inset in Fig.~\ref{Fig2} shows dc resistance $R_{xx}$  measured in sample 1 at $f_{\rm mw}=$70~GHz and $T=$2~K which displays well
resolved MIRO. Main panel in Fig.~2 presents the dependence of  $\log_{10}(\Delta R_{xx} \omega_c/\omega)$ vs. $\omega/\omega_c$,
where $\Delta R_{xx}$ is the amplitude of MIRO measured between adjacent peaks and dips. 
In accordance with Eq.~(\ref{ph}) containing $\delta^2=\exp(-2\pi/\wc\tq)$, this dependence is linear. The slope  gives the quantum
relaxation time $\tau_q=$9.1~ps which we use for calculation of QMA in sample 2.

Sample 2 was cleaved from the same wafer in vicinity to sample 1. In order to reduce the CMP effects \cite{ourPRB,Oleg,quinn76},
which can obscure weak QMA oscillations under the investigation, we cleaved a narrow 0.5 mm $\times$ 1.5 mm rectangular stripe. The
magneto-absorption experiment was performed using a home-built microwave cavity setup at liquid helium temperatures.\cite{Seck}  The cavity
with a tunable resonance frequency had a cylindrical shape with 8~mm diameter and the height between 3 and 8 mm  adjustable with a
movable plunger.  The cavity  operated in TE$_{011}$ mode, where $\{011\}$ are the numbers of half-cycle variations in the angular,
radial, and longitudinal directions, respectively. The 2DEG stripe was placed at the bottom of the cavity  with the external magnetic
field normal to the 2DEG plane and the microwave electric field of the TE$_{011}$ mode oriented along the short side of the rectangular
sample. The sample was placed in a ``face up'' fashion such that the active 2DEG layer is separated from the plunger surface by the
substrate. This geometry has higher sensitivity to weak absorption signals such as QMA but is not appropriate to study i.e. CMPs in
the vicinity of the CR due to cavity over-coupling effects. To further improve sensitivity we measure the differential signal with
respect to magnetic field.
\begin{figure}
\includegraphics[width=0.95\columnwidth]{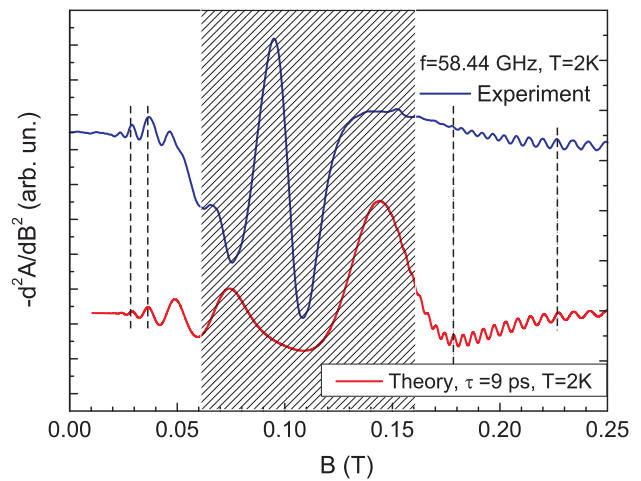}
\caption{Magneto-absorption measured on Sample~2 at $T=$2~K for $f_{\rm mw}=$58.44~GHz (top curve) and absorption coefficient ${\cal
A}$ [Eqs.~(\ref{main}) and (\ref{AD}), bottom curve] calculated without fitting parameters using $\tau_q =9.1$~ps determined from
MIRO measurements on Sample~1 (Fig.~\ref{Fig2}).
 }
 \label{Fig3}
\end{figure}

The top curve in Fig.~\ref{Fig3} presents $B$ dependence of the second derivative of the measured absorption for $f_{\rm
mw}=$58.44~GHz whereas the bottom curve shows the second $B$ derivative of the absorption coefficient ${\cal A}$, Eqs. (\ref{AD}) and
(\ref{main}), calculated without fitting parameters  using $\tau_q=9.1$~ps determined from  MIRO measurements on sample~1
(Fig.~\ref{Fig2}). Both curves demonstrate well resolved QMA with maxima at harmonics of the CR and dynamic SdHO with maxima
determined by the position of the chemical potential with respect to LLs. Clearly, the theory reproduces the experimental trace in
Fig.~\ref{Fig3} quite well everywhere except for the shaded region around the CR where the signal is distorted due to the CMP
absorption. The dimensions of sample~2 were chosen to allow only one CMP mode at $B=$0.092~T. The absence of higher modes for $f_{\rm
mw}<$63~GHz enabled the possibility to observe clear quantum oscillations on both sides of the shaded region where finite-size
effects \cite{ourPRB,Oleg,quinn76} are not essential.
 \begin{figure}
\includegraphics[width=0.95\columnwidth]{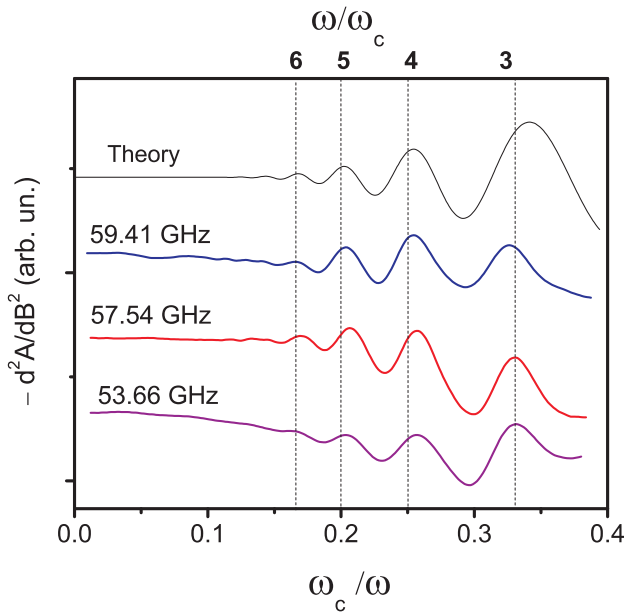}
\caption{QMA measured at different frequencies compared to the theory [Eq.~(\ref{mainOvLLs}), parameters as in Fig.~\ref{Fig3}].
 }
 \label{Fig4}
\end{figure}

The low-field traces ($\omega_c/\omega < 1/2$) of the magnetoabsorption are shown in Fig.~\ref{Fig4} for several microwave
frequencies together with the function $\delta^2 \cos(2\pi\w/\wc)$. The phase and $B$ damping of the observed QMA  follow well the
theoretical dependence (\ref{mainOvLLs}), without fitting parameters. We, therefore, believe that the observed oscillations are
indeed QMA predicted in Refs. \onlinecite{Ando} and \onlinecite{DMP03}, which provides an important experimental evidence supporting
the theory of MIRO \cite{DMP03,VA,DVAMP05} based on inter-LL transitions. In our sample, QMA are strongly damped [$\delta^2=0.02$ at
$\wc=\w/2$, $\w/2\pi=f_{\rm mw}=$58.44~GHz, and $\tau_q =9.1$~ps, see Eq.~(\ref{mainOvLLs})] which makes their observation
difficult. We expect that much stronger QMA as well as dynamic SdHO can be observed on samples with higher mobility (longer
$\tau_{\rm q}$), as simulated in Fig.~\ref{Fig1}, provided the CMP effects are avoided or sufficiently reduced.

In summary, we have observed quantum magneto-oscillations in the microwave absorption and dynamic SdHO in a high mobility 2DEG.
For this purpose we used a sensitive high-$Q$ cavity technique and developed a special setup to avoid undesirable magnetoplasmon
effects masking the quantum oscillations. Using the quantum Drude formula\cite{DMP03} and the quantum relaxation time
extracted from the MIRO measurements on the same wafer we were able to reproduce the experimental results for absorption without
fitting parameters, which provides a strong experimental support to the theory of MIRO and QMA based on inter-LL transitions.

We are thankful to A.~D.~Mirlin, D.~G.~Polyakov, B.~I.~Shklovskii, S.~A.~Vitkalov, and M.~A.~Zudov for fruitful discussions. This
work was supported by the DFG, by the DFG-CFN, by Rosnauka Grant no. 02.740.11.5072, by the RFBR, and by the NRC-CNRS project.

\end{document}